\documentstyle[12pt,epsf]{article}
\newcommand{\be}{ \begin{eqnarray}}
\newcommand{\ee}{\end{eqnarray}}
\newcommand{\beno}{ \begin{eqnarray*}}
\newcommand{\eeno}{\end{eqnarray*}}
\newcommand{\raf}[1]{(\ref{#1})}
\newcommand{\bold}[1]{\mbox{\boldmath $#1$}}    

\newcommand{\ie}{{\em i.e.}}                    
\newcommand{\MeV}{{\rm MeV}}                    
\newcommand{\rme}{{\rm e}}                      
\newcommand{\r}{{\bf r}}                        
\newcommand{\q}{{\bf q}}                        
\newcommand{\k}{{\bf k}}                        
\newcommand{\beq}{\begin{equation}}
\newcommand{\eeq}{\end{equation}}
\newcommand{\beqar}{\begin{eqnarray}}
\newcommand{\eeqar}{\end{eqnarray}}
\newcommand{\bfig}{\begin{figure}}
\newcommand{\efig}{\end{figure}}
\newcommand{\ul}[1]{\underline{#1}}             
\newcommand{\del}{\partial}                     
\newcommand{\DCC}{{\rm DCC}}                    
\newcommand{\pphi}{{\bold{\phi}}}               
\newcommand{\ppsi}{{\bold{\psi}}}               
\newcommand{\dpphi}{{\delta\bold{\phi}}}        

\begin{document}

\begin{titlepage}
\setcounter{footnote}{1}
\renewcommand{\thefootnote}{\fnsymbol{footnote}}
\noindent
{\sl Physical Review C}
\hfill
{LBNL--40175}

\hfill 
{UC--413 \ \ \ \ \ \ }

\vspace{.7cm}
\begin{center}
\ \\
{\large {\bf Dileptons from Disoriented Chiral Condensates }}
\vspace{2cm}
\ \\
{\large Yuval Kluger$^\dagger$, Volker Koch, J{\o}rgen Randrup, and Xin-Nian Wang}
\ \\
\ \\
{\it Nuclear Science Division, Lawrence Berkeley National Laboratory,\\
University of California, Berkeley, CA 94720, USA}\\
\ \\
\ \\
\vspace{2cm}
{ \bf Abstract}\\
\vspace{0.2cm}
\end{center}
\begin{quotation}
Disoriented chiral condensates or long wavelength pionic oscillations
and their interaction with the thermal environment can be a significant
source of dileptons.
We calculate the yield of such dilepton production
within the linear sigma model,
both in a quantal mean-field treatment and in a semi-classical approximation.
We then illustrate the basic features of the dilepton spectrum
in a schematic model. 
We find that dilepton yield with invariant mass near
and below $2m_{\pi}$ due to the soft pion modes
can be up to two orders of magnitude larger
than the corresponding equilibrium yield.
\end{quotation}

\vfill\noindent$^\dagger$\ \footnotesize{Current address: Theoretical Division,
Los Alamos National Laboratory, Los Alamos, NM 87545, USA.}
\renewcommand{\thefootnote}{\arabic{footnote}}
\setcounter{footnote}{0}

\end{titlepage}
 
\section{Introduction}
\label{intro}

If the quark masses are neglected in two-flavor QCD,
there would be a continuum of degenerate vacua
in which chiral symmetry is spontaneously broken.
These vacua are mutually related by $SU(2)$ rotations
and can be characterized by non-vanishing scalar (sigma), 
$\langle\sigma\rangle=\langle \bar{q} q\rangle$, and pseudo-scalar (pion),
$\langle\pi_i\rangle=\langle \bar{q}\gamma_5 \tau_i q\rangle$,
quark condensates.

In the ``Baked Alaska'' scenario
proposed by Bjorken, Kowalski, and Taylor \cite{BKT},
a space-time region protected
by a hot shell from the normal vacuum (where $\langle\pi_i\rangle=0$,
$\langle\sigma\rangle\neq 0$) can relax to a misaligned vacuum with a
non-vanishing pion condensate, or disoriented chiral condensate (DCC).
After the dispersion of the hot shell, the DCC will then decay into
normal vacuum by coherent emission of soft pions.
In relativistic heavy-ion collisions,
if the system maintains quasi-equilibrium through the chiral phase transition,
the quark masses, though small,
will prevent the correlation length from growing indefinitely.
However, recent numerical studies of the linear sigma model
\cite{rw93,GGP94,gm94,ahw,boy,kluger,kluger2,JR:PRL} have shown that
a rapid cooling-like quenching can drive the system far out of equilibrium
and lead to significant amplification of the soft pionic modes.
The resulting occupation numbers may then become large
and lead to the emission of many pions in the same isospin state.
In such an ideal scenario,
the neutral pion fraction $f$ exhibits an anomalous distribution,
$P(f)=1/2\sqrt{f}$,
which has been suggested as an experimental signal \cite{ar89,bk92,bj92}.
However,
if several separate domains are formed
(\ie\ if the size of the system is large in comparison with the correlation length),
as may well occur in heavy-ion collisions,
the signal is correspondingly degraded
and the distribution approaches its normal form,
an approximately normal distribution centered around $f$=$1\over3$
\cite{Amado96,JR:NPA}.
More advanced methods of analysis would then be needed,
such as a the use of wavelets \cite{wave} or cumulative moments \cite{minimax}.

In addition to the hadronic signals,
the electro-magnetic DCC signatures have also been also addressed
\cite{zhxw96,bvhs}. 
Since the electro-magnetic current is given by
the third component of the isovector current,
the isospin oscillation of a coherent pion field
may produce photons and dileptons
which could provide information on the early dynamical evolution of the DCC.
In Ref. \cite{zhxw96},
this emission from the oscillation of the coherent pion field
was found to be sizable only for very small invariant masses,
and thus it is hard to observe experimentally
due to the large background from $\pi_0$ Dalitz decays.
However, a DCC may also contribute to the incoherent production of dileptons.
In particular, high-momentum incoherent pions may annihilate with
the coherent DCC pions.
Since the pion phase-space density in a DCC is comparatively large
and well localized in momentum space,
this process should lead to a considerable enhancement in the dilepton spectrum
at finite invariant masses ($\simeq 2 m_\pi$)
that is rather narrow in invariant mass as well as transverse momentum.

Since the dynamical evolution leading to DCC formation is
far from equilibrium,
we shall study the electromagnetic production processes
within the framework of time-dependent field theory,
using both a quantal mean-field treatment \cite{kluger3}
and a semi-classical approximation \cite{JR:PRD}.
As these calculations treat coherent and incoherent production
on an equal footing,
the results do not depend on how that distinction is made.
Therefore,
to provide a simple understanding of the phenomenon,
including the different contributions from coherent and incoherent production,
we shall subsequently 
illustrate the essential features by means of a schematic model. 

This paper is then organized as follows.
In the next section,
we first briefly review the formulas needed
to calculate the dilepton production.
We then describe the dynamical calculations obtained with the linear sigma model
and present the corresponding results for the dilepton yields. 
Finally, we turn to the schematic model. 


 \section{Dilepton production from DCC}
 \label{dileptons}

 In general, in a non-equilibrium system, such as the dynamical
 evolution of DCC fields, the in and out states are not asymptotic 
 states and the density matrices describing the system
 are not diagonal (except in isospin and charge). In this paper,
 we will neglect quantum effect caused by the off-diagonal matrices 
 of the system in the calculation of the dilepton production.
 Therefore, the dilepton production yield can be given by the $S$ matrices
 of the electromagnetic transition between different states \cite{mclerran},
 \begin{eqnarray}\label{eq:yield0}
  \frac{dN_{\ell^+\ell^-}}{d^4q}&=&\frac{\alpha^2}{6\pi^3}\frac{B}{q^4}
    (q^\mu q^\nu-q^2g^{\mu\nu}) W_{\mu\nu}(q)\ , \nonumber \\
    W_{\mu\nu}(q)&=&\frac{1}{\cal Z}\int d^4x d^4y e^{-iq\cdot(x-y)}
    Tr [ \hat{\rho}\ \hat{j}_\mu(x)\  \hat{j}^\dagger_\nu(y) ]\ ,
 \end{eqnarray}
 where we have summed over the final states, ${\cal Z}={\rm Tr}[\hat{\rho}]$, 
 and
 \begin{eqnarray}
   B=(1-\frac{4m^2_\ell}{q^2})^{1/2}(1+\frac{2m^2_\ell}{q^2})\ .
 \end{eqnarray}
 Since we are interested in the production of electron-positron pairs,
 we shall neglect the lepton mass $m_\ell$ in the following, i.e. $B=1$.

 We have modified the temporal integration boundaries from the
 asymptotic times $t=\pm\infty$ to finite initial and final times
 $t_i$ and $t_f$. This is suitable for an initial-value problem where 
 the initial conditions are fixed in non-asymptotic states 
 which are assumed to be formed in a relativistic 
 nuclear collision.

 The Lagrangian density of the linear sigma model is given by
 \begin{equation}
   {\cal L}= \frac{1}{2}\partial_\mu \phi \partial^\mu \phi
   -\frac{1}{4}\lambda (\phi^2-v^2)^2 +H \sigma\ ,
 \end{equation}
 where $\phi=(\sigma, {\bf \bold{\pi}})$ are the chiral fields in $O(4)$
 representation. The parameters, $\lambda$, $v$ and $H$ are determined
 by the pion and sigma mass, $m_\pi$, $m_\sigma$, and the pion decay 
 constant $f_\pi$.
 In order to include electromagnetic (EM) processes,
 we can introduce an EM field in the charged sector of the above Lagrangian.
 The symmetrized EM current density coincides with the third component 
 of the isovector current density and is given by
 \begin{equation}
   \hat{j}_\mu(x)=\frac{i}{2}[\hat{\pi}^\dagger(x)
   \stackrel{\leftrightarrow}{\partial}_\mu \hat{\pi}(x)-\hat{\pi}(x)
   \stackrel{\leftrightarrow}{\partial}_\mu \hat{\pi}^\dagger(x)]
  = \hat{\pi}_1(x) \partial_\mu \hat{\pi}_2(y) - 
     \hat{\pi}_2(x) \partial_\mu \hat{\pi}_1(y)\ , 
 \end{equation}
 where the complex charged pion field operators are related to the Cartesian 
 components by  
 \begin{equation}
   \hat{\pi}(x)=\frac{1}{\sqrt{2}}[\hat{\pi}_1(x)+i\hat{\pi}_2(x)]\ , \;\;
   \hat{\pi}^\dagger(x)=\frac{1}{\sqrt{2}}[\hat{\pi}_1(x)-i\hat{\pi}_2(x)] \ .
 \end{equation}
 We shall neglect the quadratic coupling in the gauged linear sigma model
 and the anomalous electromagnetic coupling of $\pi^0$, which all contribute
 to the dilepton production only to higher orders
 in the fine structure constant $\alpha=e^2/4\pi$.

 \subsection{Mean-field treatment}
 \label{mean-field}

 In the mean-field treatment, the four-point functions 
 in the current-current correlator in Eq.~(\ref{eq:yield0}) are given 
 as products of two- and one-point functions similar to the decomposition 
 of a free field four-point function according to Wick's theorem.
 Since the one-point functions then represent the average field
 they vanish for the pions.
 For uniform density matrices the contributions from 
 two-point functions in the coincidence limit vanish since 
 $\langle\hat{j}^\mu(x)\rangle=0$. Furthermore correlators of the type 
 $\langle\hat{\pi}(x)\hat{\pi}(y)\rangle$ or 
 $\langle\hat{\pi}^\dagger(x)\hat{\pi}^\dagger(y)\rangle$ 
 contribute only to zero-momentum processes and and vanish altogether for the
 initial conditions chosen in the present calculations.
 Therefore,
 the only  remaining components for the current-current correlator
 are of the type $\langle\hat{\pi}(x)\hat{\pi}^{\dagger}(y)\rangle$.
 The current-current correlator then takes the form,

 \begin{eqnarray}\label{W}
 W_{\mu\nu}(x,y)&=&\langle\hat{\pi}^\dagger(x)\hat{\pi}(y)\rangle
   \langle\partial_\mu\hat{\pi}(x)\partial_\nu\hat{\pi}^\dagger(y)\rangle
 + \langle\partial_\mu\hat{\pi}^\dagger(x)\partial_\nu\hat{\pi}(y)\rangle
 \langle\hat{\pi}(x)\hat{\pi}^\dagger(y)\rangle 
 \nonumber \\ &-&
 \langle\partial_\mu\hat{\pi}(x)\hat{\pi}^\dagger(y)\rangle
 \langle\hat{\pi}^\dagger(x)\partial_\nu\hat{\pi}(y)\rangle - 
 \langle\partial_\mu\hat{\pi}^\dagger(x)\hat{\pi}(y)\rangle
 \langle\hat{\pi}(x)\partial_\nu\hat{\pi^\dagger}(y)\rangle.
 \label{eq:yield01}
 \end{eqnarray}

 For a system in thermal equilibrium at the temperature $T$,
 the ensemble average is
 $\langle \cdot\rangle_{th}={\rm Tr}[\hat{\rho}_{th}\cdot]$,
 where $\hat{\rho}_{th}\sim {\rm e}^{-\hat{H}/T}$.
 For free pions,
 the eigenstates are plane waves and the thermal ensemble average can be
 expressed accordingly,
 \begin{eqnarray}
   \langle\hat{\pi}^\dagger(x) \hat{\pi}(y)\rangle_{th}&=&
   \int\frac{d^3k}{2\omega_k (2\pi)^3} [ n^+_k e^{ik\cdot(x-y)}
   +(1+n^-_k) e^{-ik\cdot(x-y)}]\ , \nonumber \\
   \langle\hat{\pi}(x) \hat{\pi}^\dagger(y)\rangle_{th}&=&
   \int\frac{d^3k}{2\omega_k (2\pi)^3}[ n^-_k e^{ik\cdot(x-y)}
   +(1+n^+_k) e^{-ik\cdot(x-y)}]\ , \label{eq:thprop}
 \end{eqnarray}
 where $n^\pm_k=1/(e^{\omega/T}-1)$
 are the Bose-Einstein occupation numbers for the charged pions.

 The hadronic tensor (\ref{W}) entering into the dilepton-production expression
 then becomes
 \begin{eqnarray}
   \frac{W^{th}_{\mu\nu}(q)}{d^4x}&=&\int \frac{d^3k_1}{2\omega_1(2\pi)^3}
   \frac{d^3k_2}{2\omega_2(2\pi)^3} (2\pi)^4  \left\{ 
   (k_1-k_2)_\mu(k_1-k_2)_\nu \right. \nonumber \\ 
   &\cdot& [n^-_{k_1}n^+_{k_2} \delta^4(q-k_1-k_2)+
   (1+n^-_{k_1})(1+n^+_{k_2})\delta^4(q+k_1+k_2)] \nonumber \\
   &+& \left. (k_1+k_2)_\mu(k_1+k_2)_\nu [ n^-_{k_1}(1+n^-_{k_2}) 
   +n^+_{k_1}(1+n^+_{k_2})]\delta^4(q-k_1+k_2) \right\}\; , 
 \label{eq:yield5}
 \end{eqnarray}
 where we have taken advantage of the translational invariance
 in space and time to obtain the ensemble average {\em rate} of production
 by dividing by the corresponding four-volume $V(t_f-t_i)$.

 The three terms in the above relation correspond to pion pair
 annihilation, creation and bremsstrahlung, respectively.
 The factor $1+n^\pm_k$ is a result of the Bose
 enhancement in the final states. 
 For dilepton production ($q^2>0$, $q_0>0$) only the pair annihilation
 term remains due to energy and momentum conservation. However, as we shall
 discuss briefly in section \ref{schematic}, once the pion dispersion relation
 develops several branches due to interactions,  one can simply replace  
 $n^\pm_k$ $(1+n^\pm_k)$ by $\sum_i n^\pm_k(i)$ [$\sum_i(1+n^\pm_k(i))$] in 
 the above equation with summation
 over the number of branches in the dispersion relation. Since dileptons
 can be emitted via the transition from one branch to another,
 all the terms, except the pair creation process, are allowed.

 When the system is not in equilibrium,
 as is generally the case in a collision scenario,
 the current-current correlator cannot be expressed explicitly.
 However,
 in the present mean-field approximation the pion field vanishes on the average,
 and so the pion field operators can be expanded
 in terms of the annihilation and creation operators
 for medium-modified charged pions,
 \begin{equation}
 \hat{\pi}(x,t)\equiv \int \frac{d^3k}{(2\pi)^3}
  e^{i{\bf{k}} \cdot{\bf{x}}} [ f_k(t)\hat{a}_k 
  + f_k^{\ast}(t)\hat{b}_{-k}^{\dagger}]\ .
 \label{fieldexpB}
 \end{equation}
 where the coefficients are the time-dependent mode functions $f_k(t)$.

 The time evolution of the mode functions can be obtained in
 the mean-field approximation 
 and we can then obtain the two-point correlation functions in 
 Eq.~(\ref{eq:yield01}), as described in Ref.~\cite{kluger,kluger3}.
 To solve the time evolution 
 of the mode functions one needs to specify the initial
 conditions of the mean-field $\langle\sigma\rangle$
 and its time derivative, the mode functions and their time
 derivatives, the average occupation numbers for each mode
 $\langle \hat{a}_k^\dagger \hat{a}_k\rangle\equiv n^+_k$,
 $\langle \hat{b}_k^\dagger \hat{b}_k\rangle\equiv n^-_k$ and
 their pair correlation $\langle\hat{a}_k\hat{b}_{-k}\rangle\equiv F_k$.

 In order to have a finite set of renormalized equations, we have to 
 choose the mode functions so that the high-momentum modes coincide 
 with the zeroth order adiabatic vacuum described by
 \begin{equation}
 f_k(t_0)={1\over\sqrt{ 2\omega_k(t_0)}}, \,\,
 \dot{f}_k(t_0)=
 \left[-i\omega_k(t_0)-{\dot{\omega}_k(t_0)\over
 2\omega_k(t_0)}\right]f_k(t_0)\ , 
 \label{initial_modes}
 \end{equation}
 with $\omega_k^2(t_0)=k^2+<\chi(t_0)>$ and 
 $\chi=\lambda(\hat{\phi}^2/8-v^2)$.

 In this study,
 we will select ensembles with $n^+_k=n^-_k$
 and vanishing pair correlations, $F_k=0$.
 By calculating two- and four-point functions the parameters 
 $\lambda$, $H$, and $v$ were chosen to give the best fit to the
 physical observables $f_\pi$, $m_\pi$, and the $\delta^{I=0}_s$ 
 phase shifts \cite{kluger}.
 In these simulations the bare coupling constant is $\lambda=20$, 
 the momentum cutoff is $\Lambda= 1 \, \rm GeV $,
 the pion mass is $135 \, \rm MeV$
 and the pion decay constant is $f_{\pi}=92.5 \, \rm MeV$.
 The resulting mass of the $\sigma$ is $340 \, \rm MeV$.

 We investigate two different initial ensembles and evolve
 them from time $t_i=-50$ fm/$c$ to $t_f=50$ fm/$c$.
 All cases were prepared with equivalent energy densities
 and with zero kinetic energy density associated with the $\sigma$ mean-field.
 The vacuum adiabatic mode functions are assumed as 
 \begin{equation}
 f_k(t_0)={e^{-i\omega_k t_0}\over \sqrt{2\omega_k}}\ ,\
 \dot{f}_k(t_0)=-i\omega_k f_k(t_0)\ .
 \end{equation}

 The first one is an ensemble which at the initial time ($t=-50$~fm/$c$) 
 has a mean $\sigma$ field which is slightly perturbed around
 a thermal ensemble prepared at $T$=100~MeV.
 The initial quasiparticle frequency in this case is given by
 \begin{equation}
 \omega^2_k=k^2+<\chi>\ ,
 \end{equation}
 with the initial value $<\chi>= (138 {\rm MeV})^2$. 
 The initial value of the mean $\sigma$ field is 87~MeV.
 The average quasiparticle population $n_k$ is given by
 the corresponding Bose-Einstein distribution and
 the energy density is then $\epsilon= 13.7 \, \rm MeV/fm^3$. 
 At this relatively low temperature,
 the ensemble should be close to that of free pions
 and the results may thus serve to assess the quality of the calculation.
 Subsequently, 
 we will call these initial conditions {\em thermal}.

 The second ensemble is prepared with a negative value of $\chi$
 in order to emulate a quench scenarios.
 In order to initialize the unstable pion modes
 (those having a negative value of $\omega_k^2$),
 we have employed the relation
 $\omega^2_k=k^2+m_{\pi}^2+(<\chi>-m_{\pi}^2)\exp{(-f_{\pi}^4/ k^4)}$,
 using the initial value $<\chi>=- (132 \, \rm MeV)^2$.
 [We note that at a large momentum
 the frequency agrees with the adiabatic vacuum frequency,
 as required by renormalization.]
 The initial value of the mean field is $<\sigma> = 31 \, \rm MeV$
 and the energy density is $\epsilon=13.6 \, \rm MeV/fm^3$,
 \ie\ similar to the equilibrium scenario.
 In the quench case,
 the initial quasiparticle modes are unoccupied, $n_k=0$, 
 corresponding to the local vacuum.

\begin{figure}[htb]
 \setlength{\epsfxsize=0.7\textwidth}
 \centerline{\hspace{0.15\textwidth}\epsffile{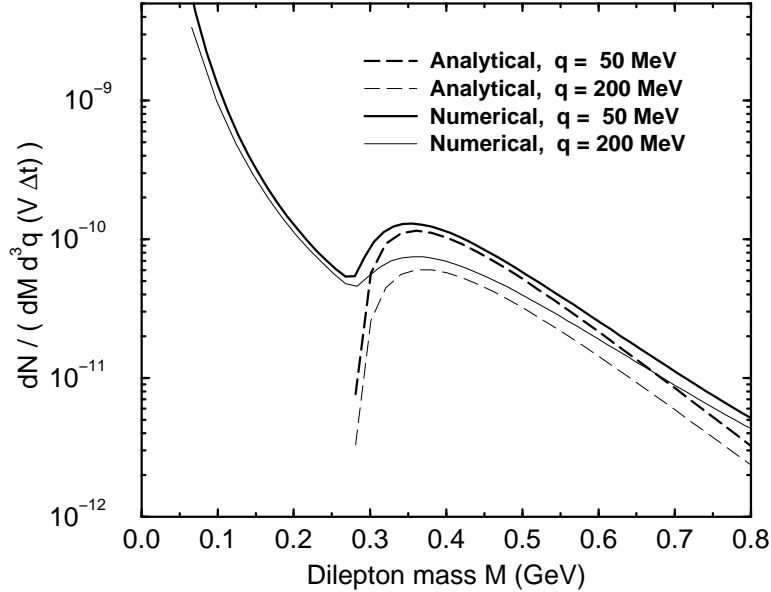}}
 \caption{Dilepton invariant mass spectrum for the thermal initial conditions
 for two values of the three momentum $q$ (full lines). Also shown are the
 theoretical results for a free pion gas (dashed lines).}
 \label{fig:theory_comp}
 \end{figure}

 Let us now turn to the results for dilepton production.
 In Fig.~\ref{fig:theory_comp} we show
 the dilepton yield obtained by using the mode functions generated from
 the thermal initial conditions to calculate the current-current correlation 
 function in Eq.~(\ref{eq:yield01}) (full lines). Note that we have divided the
 yield by the total space-time volume, $V \Delta t$, it is therefore the
 averaged production rate during the time interval $\Delta t=t_f-t_i$.
 We show the results for two 
 different values of the three momentum of the dilepton, 
 $q = 50 \, \rm MeV$ and $q = 200 \, \rm MeV$. This is compared 
 with the analytical result for a free-pion gas (dashed lines),
 \be
   \frac{1}{V T} \frac{dN_{\ell^+\ell^-}^{(2)}}{dM d^3q}
   =\frac{\alpha^2 }{48 \pi^4} \frac{q_0}{M} (1-\frac{4m_\pi^2}{M^2})
   \int_{\omega^-}^{\omega^+} \frac{d\omega}{q} n^+_k n^-_{q-k} \;\; ,
   \label{eq:theory}
 \ee
 with $\omega^\pm=(q_0 \pm q \sqrt{1-4m_\pi^2/M^2})/2$ and 
 $q_0 = \sqrt{M^2 + q^2}$. 
 As compared to the analytical result for the free pion gas,
 the numerical calculation based on the thermal mode functions
 give a contribution below the two-pion threshold as well.
 This contribution is spurious
 as it arises from the finite resolution
 of the energy $\delta$ function in our calculation:
 Since the Fourier transform in time extents only over a finite time interval, 
 the energy is not perfectly conserved. As a result the terms proportional to 
 $n_{k_1}(1 + n_{k_2})$ in Eq.~(\ref{eq:yield5}) do contribute to the 
 time-like sector while with perfect energy conservation their 
 contribution is restricted to space-like photons only. 
 This spill-over from the space-like region is amplified because of the factor
 $1/q^4$ in Eq.\ \raf{eq:yield0}.
 In addition to leading order this contribution depends on linearly on the 
 distribution 
 function $n_k$ and is thus further enhanced by $1/n_k \gg 1$ with respect
 to the pion annihilation term.   

\begin{figure}[htb]
 \setlength{\epsfxsize=0.7\textwidth}
 \centerline{\hspace{0.15\textwidth} 
 \epsffile{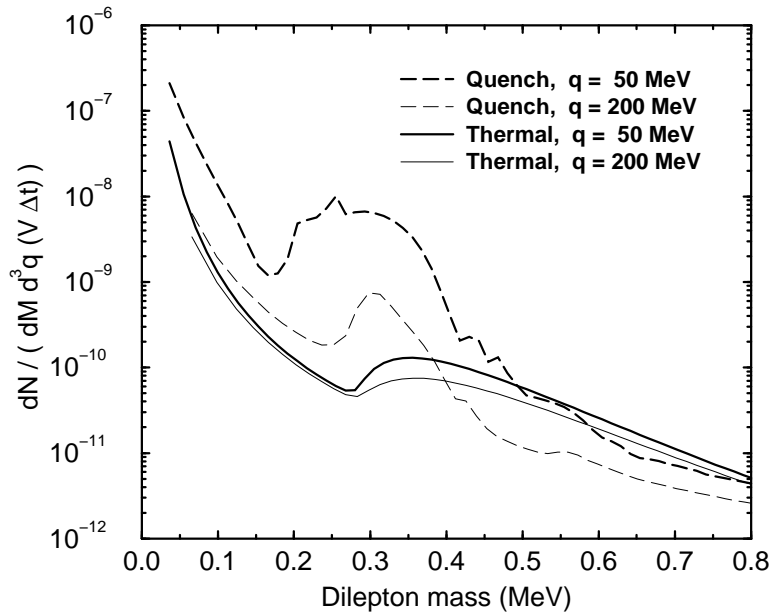}}
 \caption{Dilepton invariant mass spectra for thermal (full lines) and quench
 (dashed lines) initial conditions. Shown are the spectra for two different
 values of the three dilepton three momentum $q$.}
 \label{fig:quench_comp}
 \end{figure}

\begin{figure}[htb]
 \setlength{\epsfxsize=0.7\textwidth}
 \centerline{\hspace{0.15\textwidth} 
 \epsffile{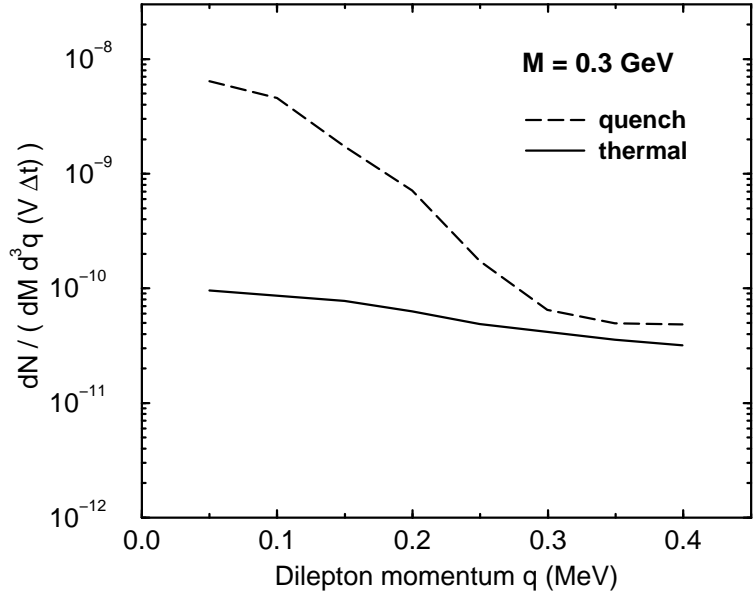}}
 \caption{Momentum spectra for thermal (full lines) and quench
 (dashed lines) initial conditions for dilepton pairs of invariant mass
 $M= 300 \, \rm MeV$.} 
 \label{fig:qplot}
 \end{figure}

 In order to verify this,
 we have also performed a calculation with the mode functions of a free pion 
 gas.
 The result is indistinguishable from the one obtained
 with the dynamical mode functions,
 showing that the interaction plays a negligible role at $T$=100~MeV.
 By inspecting the individual contributions,
 we have also verified that the spurious contribution is indeed
 due to the terms proportional to $n_{k_1} (1 + n_{k_2})$. 
 At the present stage,
 this spurious contribution forms an numerical background.
 It is also responsible for the slight disagreement between the analytical
 and numerical results for the free pions above the two-pion threshold. 
 With this numerical background in mind, let us turn now to the comparison
 between the thermal and quench initial conditions,
 as illustrated in Fig.\ \ref{fig:quench_comp}.
 Clearly,
 around an invariant mass of $M\simeq 2 m_\pi$
 the dilepton production rate from the quench initial conditions
 is about two orders of magnitude larger than the thermal production.
 This enhancement is less for larger momenta,
 reflecting the narrow momentum distribution of pions in the DCC fields.
 Furthermore,
 at low invariant masses there seems to be an enhancement
 as predicted in Ref.\ \cite{zhxw96}.
 Because of our numerical background,
 we are not able to give a quantitative estimate of this enhancement. 
 However, due to the strong background from the $\pi_0$ Dalitz decay,
 this enhancement is difficult to measure experimentally.
 Therefore, let us concentrate on the enhancement around $M \simeq 2 m_\pi$.
 As expected from the narrow momentum distribution of the DCC pions,
 this enhancement is localized in  invariant mass
 as well as in transverse momentum.
 The latter can be seen from Fig.\ \ref{fig:qplot}
 where the dilepton yield is plotted
 as a function of the three-momentum of the dilepton for
 an invariant mass of $M = 300 \, \rm MeV$.
 The enhancement is confined to low momenta, below $q \simeq 300 \, \rm MeV$.   
 While the localization of the enhancement in invariant mass
 nicely reflects the enhancement of low-momentum pion modes,
 it may be very difficult to observe experimentally.
 At low transverse momentum the background from false pairs is largest
 and it remains to be seen if present detector designs allow
 a sufficiently accurate subtraction
 to make the extraction of this signal feasible.
 In order to get an idea about the experimental constraints,
 we have subjected our results to the CERES acceptance cuts \cite{CERES}. 
 The resulting invariant mass spectrum
 is virtually indistinguishable from the thermal one.
 Only if the momentum cuts can be pushed down to $\simeq 100 \, \rm MeV$
 will an enhancement of about a factor of ten remain.

\subsection{Semi-classical treatment}
\label{semi-class}

In order to verify that our results are robust with respect to model details,
we have also calculated dilepton production with a different treatment
of the linear sigma model,
namely the semi-classical approach described in Ref.\ \cite{JR:PRD}.
In that treatment,
the system is described by the real field $\pphi(\r,t)=(\sigma,\bold{\pi})$,
which is evolved by the classical equation of motion,
\beq
[\Box +\lambda(\phi^2-v^2)]\pphi=H\hat{\sigma}\ ,
\eeq
where $\phi^2=\pphi\circ\pphi$
and $\hat{\sigma}$ denotes a unit vector along the $\sigma$ axis.
In the present context,
when we are concerned with dilepton production
in a macroscopically uniform system,
it is natural to enclose the system in a (sufficiently large) box
with periodic boundary conditions.
The spatial average of the field can then be considered
as the order parameter,
$\ul{\pphi}(t)$$=$$<$$\pphi(\r,t)$$>$,
and the residual fluctuations
represent quasi-particle excitations relative to that constant field,
$\dpphi(\r,t)$$=$$\pphi(\r,t)-\ul{\pphi}(t)$.
A similar separation holds for the time derivative of the field.

In order to solve the above equation of motion,
it is necessary to specify the initial field configuration,
$\pphi(\r,t_i)$,
and the associated time derivative.
Employing the method developed in Ref.\ \cite{JR:PRD},
we sample these initial conditions from a thermal ensemble
at a specified temperature $T$.
The occupation numbers $n_\k$ of the quasi-particle modes
are drawn from the appropriate Bose-Einstein distributions
which ensures that the initial state displays the proper quantum-statistical
features.
The resulting treatment is then akin to the Vlasov model
often employed in nuclear dynamics at moderate energies,
which is the semi-classical analogue of the
Time-Dependent Hartree-Fock description.

In order to make contact with the quantum mean-field studies described earlier,
we consider two scenarios having the same energy,
namely thermal equilibrium at $T$=140~MeV
and a corresponding quenched initial condition.
The quenched ensemble has been generated from the thermal ensemble
by suppressing the field fluctuations by a factor of 100
and resetting the order parameter to $\ul{\pphi}(t_i)=(\sigma_0,\bold{0})$,
with a vanishing time derivative, $\ul{\ppsi}(\r,t_i)=(0,\bold{0})$.
Using $\sigma_0=32\ \MeV$ ensures that the energy of the quenched scenario
matches the thermal value, which is about 40 MeV/fm$^3$.

The numerical solution of the equations of motion of the chiral fields
yields the evolution of the Cartesian components of the pion field,
$\bold{\pi}(\r,t)$,
in addition to the sigma field $\sigma(\r,t)$.
The corresponding electromagnetic current density is then easy to extract,
\beq
J_\mu(x)\ = \pi_1(x)\del_\mu\pi_2(x) - \pi_2(x)\del_\mu\pi_1(x)\ .
\eeq
In order to calculate the associated invariant differential dilepton yield,
we employ the following expression \cite{mclerran},
\beqar\label{d4N}
{d^4N\over d^4q}\ &=&\ {2\over3\pi} ({\alpha\over2\pi})^2
\left( {q^\mu q^\nu \over q^4} - {g^{\mu\nu} \over q^2} \right)
\int d^4x \int d^4y\ J_\mu(x)\ \rme^{-iq(x-y)}\ J_\nu(y)\\ \label{JJ}
&=&\ {2\over3\pi} ({\alpha\over2\pi})^2\ \tilde{J}^*_\mu(q)
        \left( {q^\mu q^\nu \over q^4} - {g^{\mu\nu} \over q^2} \right)
        \tilde{J}_\nu(q)\ .
\eeqar
This expression ignores the final-state Bose enhancement factors 1+$n_\k$
which is justified when the occupation number $n_\k$ is small,
as is typically the case in equilibrium \cite{JR:PRD}.
In particular,
for $T$=140~MeV the occupancy of the lowest quasi-particle mode is about 0.30.
However,
the occupancy  may reach several units in the corresponding quench scenario
and the formula (\ref{d4N}) may then significantly underestimate
the contribution from the softest pion modes\footnote{
It may be noted that if the calculated pion field $\bold{\pi}(\r,t)$
is assumed to represent a standard coherent state,
then the quantal evaluation of the dilepton radiation rate
would lead to the above expression (\ref{d4N})
when the commutator terms are ignored;
if those commutator terms were retained,
then the final-state Bose enhancement factors would be recovered.}.

The last expression (\ref{JJ}) recasts the dilepton yield
in terms of the four-dimensional Fourier transform of the charge current,
\beq\label{Jq}
\tilde{J}_\mu(q)\ =\ \int d^4x\ J_\mu(x)\ \rme^{iqx}\
=\ \int_{t_i}^{t_f} dt\int_V d\r\ J_\mu(\r,t)\ \rme^{iq_0 t-i\q\cdot\r}\ .
\eeq
The factorized form (\ref{JJ}) is of great numerical convenience,
since the transform $\tilde{J}_\mu(q)$ can be readily accumulated
in the course of each dynamical history,
and the contraction in (\ref{JJ}) need only be carried out
at the end of the evolution.
In (\ref{d4N}) and (\ref{Jq}) the time integration extends
over the duration of the observation,
from $t_i$ to $t_f$.
In the equilibrium scenario,
the yield then becomes proportional to $\Delta t=t_f-t_i$
as well as to the volume $V$,
and so a division by the four-volume $V \Delta t$ yields the corresponding
invariant production rate $d^4N/(d^4q \,(V \Delta t))$.

In general,
we consider an entire sample of $\cal N$ individual evolutions,
$\{\pphi^{(n)}(\r,t)\}$,
where the label $n$ enumerates the individual ``events'' in the sample.
The resulting ensemble-average dilepton yield is then
\beq
\prec {d^4N\over d^4q}\succ\ =\
{1\over\cal N}\sum_{n=1}^{\cal N} {d^4N^{(n)}\over d^4q}\ ,
\eeq
where ${d^4N^{(n)}/d^4q}$ is the contribution from the particular event $n$,
obtained as described above.
Since we consider ensembles that have translational symmetry,
the current-current correlation function, $\prec J_\mu(x)\ J_\nu(y)\succ$,
will depend only on the spatial separation.
Moreover,
in the special case of an equilibrium ensemble,
its temporal dependence in equilibrium is only via the time difference.
[In practice,
the considered dilepton observables vary little from  event to event,
because the system is larger than the correlation length,
and therefore sufficiently accurate results
can be obtained on the basis of rather small samples.]

In order to verify that the adopted method indeed leads to
physically reasonable results,
let us consider the production of back-to-back dileptons
from a thermal gas of free pions.
In that special case,
the four-momentum of the dilepton is of the form $q=(M,\bold{0})$
and we are interested in masses $M$ above $2m_\pi$.
The current-current contraction in (\ref{d4N}) is then especially simple
and it is elementary to show that its ensemble average is given by
\beq
\prec J_\mu(x)
\left({q^\mu q^\nu \over q^2}-{g^{\mu\nu}}\right)
J_\nu(y)\succ\ =\ 
\prec {\bold{J}(x)\cdot\bold{J}(y)} \succ\ =\
2|{\nabla}C|^2-2C\Delta C\ ,
\eeq
where we have employed the thermal correlation function
of the charged pion fields,
\beq
\prec\pi_1(x)\pi_1(y)\succ\ =\
\prec\pi_2(x)\pi_2(y)\succ\ =\
C(\r,t)\ =\ {1\over V}\sum_\k {\tilde{n}_k \over \omega_k}\
\cos(\k\cdot\r-\omega_k t)\ ,
\eeq
with $\tilde{n}_k$ being the thermal occupancy,
$\tilde{n}_k=1/(\exp(\omega_k/T)-1)$,
and $(\r,t)$ denoting the difference $x$-$y$.
[The adopted sampling procedure ensures that the numerically extracted
correlation function indeed yields this expression \cite{JR:PRD}.]
Since the back-to-back dileptons have vanishing momentum, $\q=\bold{0}$,
the Fourier transform over the separation $\r$ reduces to a spatial average
and we readily find
\beq
\int_V d\r\ \prec \bold{J}(x)\cdot\bold{J}(y)\succ\ =\
4\int_V d\r\ |\nabla C(\r,t)|^2\ =\ {2\over V} \sum_\k
{\tilde{n}_k^2 \over \omega_k^2} {k^2}\ [1+\cos2\omega_kt]\ .
\eeq
The remaining Fourier transformation over the temporal difference
then restricts the contributions in the sum to those modes
that have frequencies $\omega_k$ near half the dilepton mass, $M/2$.
Thus, in the continuum limit,
when both the box and the time interval are large,
we recover exactly the usual expression for production of dileptons
by pion annihilation,
\beqar
{d^4N^{th}\over d^4q d^4x} =
{4\over3\pi} ({\alpha\over2\pi})^2 \int {d\k\over (2\pi)^3}\int dt\
{\tilde{n}_k^2\over\omega_k^2}{k^2\over M^2}\ [1+\cos2\omega_kt]\ {\rm e}^{iMt}
= {\alpha^2\over3}{n_0^2\over(2\pi)^4}
\left(1-{4m_\pi^2\over M^2}\right)^{3\over2} ,
\eeqar
where $n_0$ denotes the occupancy of pion states with the matching frequency
$\omega_0=M/2$.
Thus, at the formal level,
the semi-classical method is indeed physically reasonable. 

\begin{figure}[htb]
\setlength{\epsfxsize=0.7\textwidth}
\centerline{\hspace{0.15\textwidth}
\epsffile{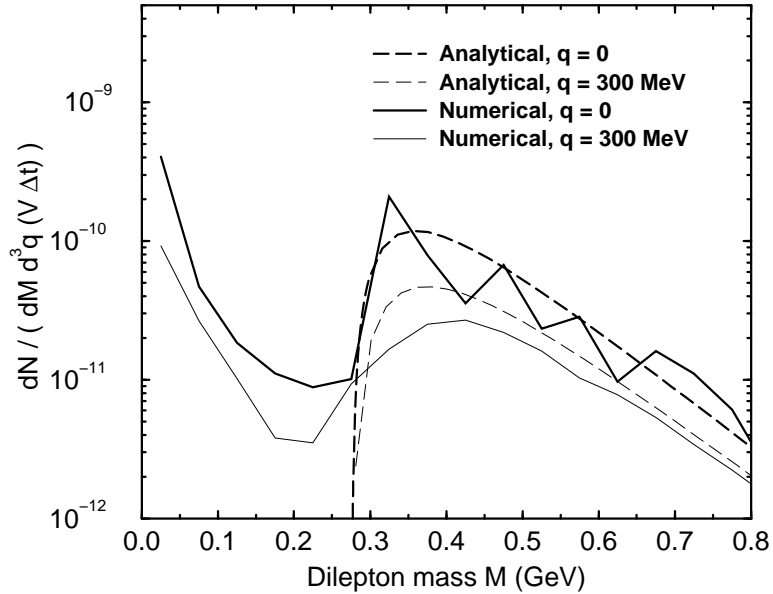}}
\caption{The invariant dilepton production rate
in a gas of free pions in thermal equilibrium at $T=100\ \MeV$.
The solid curves are obtained by solving the free-field equations numerically,
while the dashed curves represent the corresponding analytical results.
The heavy curves are for back-to-back production, $q=0$,
while the light curves are for a finite momentum of the dilepton,
$q=300\ \MeV$.}
\label{JR:1}
\end{figure}

In order to illustrate the numerical reliability of the calculations,
we consider again the simple case of a free pion gas
in thermal equilibrium,
in analogy with what was shown in Fig.\ \ref{fig:theory_comp}.
Figure \ref{JR:1} shows 
both the analytical result for the invariant dilepton production rate
and the corresponding results obtained by integrating
the equation of motion for the free fields,
with the initial field configurations having been sampled
from the associated thermal ensemble as described above.
As in the case of the mean-field treatment,
the numerical calculation yields a reasonable result
for dilepton masses above the pion annihilation threshold,
but exhibit a divergent behavior for lower masses
as a reflection of the pole at $q=0$.
The fluctuations in the numerical results
are a consequence of the much coarser grid employed in the semi-classical
calculations:
the system is confined within a cubic torus of size $L$=24~fm
(with a grid spacing of 0.24~fm)
and 41 modes have been included in each of the three Cartesian directions,
for a total of 68,921 modes.
Moreover, the time evolution has been performed only up to 50 fm/$c$.

\begin{figure}[htb]
\setlength{\epsfxsize=0.7 \textwidth}
\centerline{\hspace{0.15\textwidth}
\epsffile{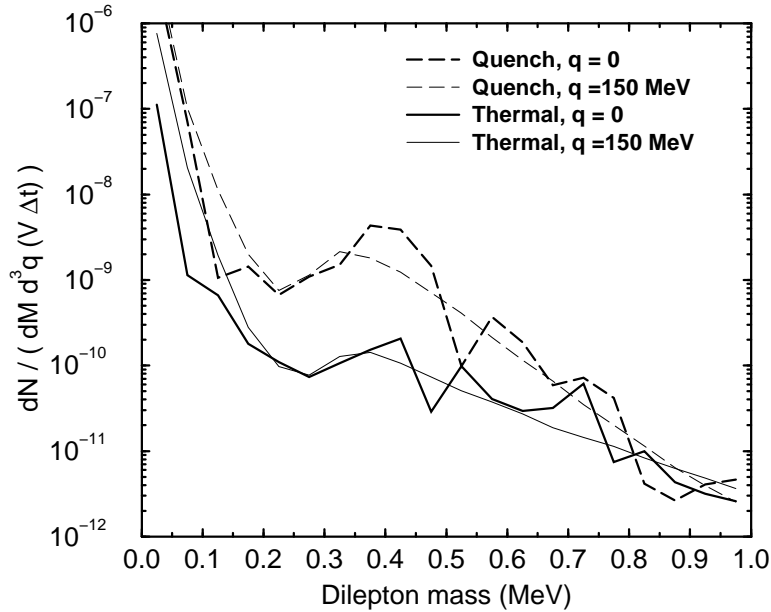}}
\caption{The invariant mass spectrum for both thermal initial conditions
(solid curves) and the corresponding quench scenario (dashed curves),
for either back-to-back dileptons having {\bf q}={\bf 0} (heavy curves)
and dileptons with a finite momentum of $q$=150~MeV(light curves).}
\label{JR:2}
\end{figure}

We now move on to discuss the results of the numerical simulations
in the two scenarios described above.
Figure \ref{JR:2} shows the invariant production rate
as a function of the dilepton mass $M=\sqrt{q^2}$,
for various magnitudes of its momentum $\q$,
in a display similar to Fig.\ \ref{fig:theory_comp}.
The results are qualitatively similar to those obtained with the
mean-field treatment:
the quench scenario leads to a large enhancement
around 400 MeV.
In the present case the enhancement is only about one order of magnitude,
because the coarser grid considered leads to a quicker damping of the
oscillations in the order parameter and, consequently,
the amplification of the soft pions modes is less extreme.
Still, there is clearly a very significant effect of the
non-equilibrium evolution.

\begin{figure}[htb]
\setlength{\epsfxsize=0.7 \textwidth}
\centerline{\hspace{0.15\textwidth}
\epsffile{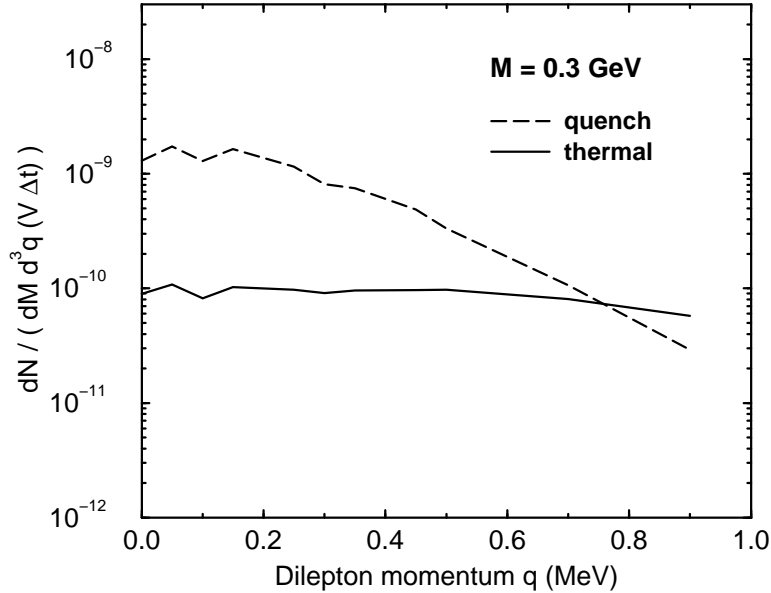}}
\caption{
The production rate $(d^4N/dM d^3q \,(V \Delta t))$
as a function of the magnitude of the dilepton momentum $\q$,
for dilepton masses near $M$=300~MeV.}
\label{JR:3}
\end{figure}

Finally,
Fig.\ \ref{JR:3} shows the effective production rate 
$d^4N/(dM d^3q \,(V \Delta t))$  
as a function of the magnitude of the dilepton momentum $\q$,
for dilepton masses near $M$=300~MeV.
Again,
we see how the non-equilibrium evolution following the quench
caused a large enhancement of the slow-moving dileptons.
The  results of the two calculations agree qualitatively: The quench initial
conditions lead to a substantial enhancement of the dilepton yield at invariant
masses close to $M \simeq 2 m_\pi$. The calculation based on the
quantal mean-field approximation (section \ref{mean-field}) predicts a
considerably stronger enhancement which is also more narrow in invariant mass
as well as momentum. As we will explain in the following section this
implies that in the quantal mean-field approximation the enhancement of the low
momentum modes is stronger and narrower in momentum space than in the
semi-classical approximation. Translated into the DCC language, the DCC fields
generated in the quantal mean-field approximation have a larger field strength
and are of larger spatial extent. 
There are several possible reasons for this quantitative difference. 
First,
the Bose enhancement factors, present in the mean-field treatment,
certainly  will increase the dilepton yield to some extent.
Second, the comparatively coarse grid employed in the semi-classical treatment
leads to a faster dampening of the oscillations in the order parameter,
resulting in a smaller amplification of the soft pion modes.
Finally, the semi-classical treatment incorporates the mode mixing
arising from the non-linear form of the interaction.
This mechanism helps to equilibrate the system 
and thus redistribute the strength gained by the low-momentum modes. 
While it would be interesting to see if the inclusion of mode mixing
in the quantal mean-field formulation reduces the enhancement,
it will be hard to give a precise quantitative estimate of their effect.
Therefore,
it is probably more reasonable to consider the quantitative difference between 
the two approximations as an inherent uncertainty which can only be resolved 
by experiment.
We note, however, that the above difference clearly demonstrates that a precise
measurement of the dilepton spectrum not only can reveal the presence of DCC
configurations but can also provide information about their size as well as
their strength.


\section{A schematic model}
\label{schematic}

With the mean-field and semi-classical treatments,
the time-dependent fields are dynamically coupled to the quasi-particle modes
and the soft and hard modes of the fields are treated on an equal footing.
To understand the main features  of dilepton production due to the soft modes,
we shall now consider the dilepton production process
in a schematic model in which the soft and hard modes
are coupled only via the electromagnetic interaction
causing dilepton production.
The hard modes are then represented by a thermal gas of pions
having a specified temperature $T$,
while the soft modes will be evolved dynamically
according to the  one-dimensional Bjorken expansion scenario.
Hence the density operator can be factorized,
$\hat{\rho}=\hat{\rho}_{th} \otimes \hat{\rho}_{DCC}$,
where $\hat{\rho}_{th}$ is the normalized density operator for the thermal gas
and $\hat{\rho}_{DCC}$ is the normalized density operator for the soft modes,
referred to as the \DCC\ field.

The current-current correlator $\hat{W}_{\mu\nu}(q)$ can then be
decomposed according to how many thermal pions are involved
in the dilepton production process.
The first term represents the coherent emission of dileptons from \DCC\ fields
alone and involves no thermal pions,
\begin{equation}
  W^{(0)}_{\mu\nu}(q)\ =\ \int d^4x d^4y\
  \langle\hat{j}_\mu(x)\hat{j}^\dagger_\nu(y)\rangle_{DCC}\
        {\rm e}^{-iq\cdot(x-y)}\
  =\ \langle\hat{j}_\mu(q)\hat{j}^\dagger_\nu(q)\rangle_{DCC}\ .
  \label{eq:yield1}
\end{equation}
where the \DCC\ expectation value is
$  \langle \cdots\rangle_{DCC}={\rm Tr}[\hat{\rho}_{DCC}\cdots ]$.

Contributions to the current-current correlator
from processes involving a thermal pion are
\begin{eqnarray}
  W^{(1)}_{\mu\nu}(x,y)&=&\langle\hat{\pi}^\dagger(x)\hat{\pi}(y)\rangle_{DCC}
  \langle\partial_\mu\hat{\pi}(x)\partial_\nu\hat{\pi}^\dagger(y)\rangle_{th}
  -\langle\partial_\mu\hat{\pi}^\dagger(x)\partial_\nu\hat{\pi}(y)\rangle_{DCC}
\langle\hat{\pi}(x)\hat{\pi}^\dagger(y)\rangle_{th}
      \nonumber \\ &+&
\langle\partial_\mu\hat{\pi}^\dagger(x)\hat{\pi}(y)\rangle_{DCC}
\langle\hat{\pi}(x)\partial_\nu\hat{\pi^\dagger}(y)\rangle_{th}
-\langle\partial_\mu\hat{\pi}(x)\hat{\pi}^\dagger(y)\rangle_{DCC}
      \langle\hat{\pi}^\dagger(x)\partial_\nu\hat{\pi}(y)\rangle_{th}
\nonumber \\  &+& c.c. \;\; , \label{eq:yield2}
\end{eqnarray}
where we have omitted the vanishing disconnected part and all other terms 
containing $\langle \hat{\pi}^\dagger(x) \hat{\pi}^\dagger(y)\rangle_{th}$
or $\langle \hat{\pi}(x) \hat{\pi}(y)\rangle_{th}$ which vanish
in a thermal equilibrium environment.
Using Eq.~(\ref{eq:thprop}), we obtain
\begin{eqnarray}
  W^{(1)}_{\mu\nu}(q)&=&\int\frac{d^3k}{2\omega_k (2\pi)^3}
  \left\{ (2k_\mu+q_\mu)(2k_\nu+q_\nu) 
  [(1+n^+_k) \langle\hat{\pi}^\dagger(-q-k)\hat{\pi}(-q-k)\rangle_{DCC}
  \right. \nonumber \\
  &+&(1+n^-_k) \langle\hat{\pi}(q+k)\hat{\pi}^\dagger(q+k)\rangle_{DCC}]
  +(2k_\mu-q_\mu)(2k_\nu-q_\nu) \nonumber \\
  &\cdot& \left.
  [ n^+_k \langle\hat{\pi}(q-k)\hat{\pi}^\dagger(q-k)\rangle_{DCC}
  +n^-_k \langle\hat{\pi}^\dagger(k-q)\hat{\pi}(k-q)\rangle_{DCC}]
  \right\}, \label{eq:yield3}
\end{eqnarray}
The first two terms in the above equation correspond
to the emission of one pion together with a dilepton by the \DCC\ field.
The factor $1+n^\pm_k$ is a result of the Bose
enhancement in the final state.
The second  two terms proportional to $n^\pm_k$ represent the annihilation
or absorption of one thermal pion by the \DCC\ field.

{}From Eq.\ (\ref{eq:yield3}) one can readily see
how the momentum distribution of the \DCC\ field is imprinted
onto the dilepton spectrum.
Once the momentum of the dilepton is large
compared to the inverse of the \DCC\ domain size, 
the integral no longer has support from the 
Fourier transform of the pion field 
$\langle\hat{\pi}^\dagger(k\pm q)\hat{\pi}(k\pm q)\rangle_{DCC}$,
which restricts the contribution to low momenta and to a small window
in invariant mass.
For example, if one considers a classical field that oscillates
with a typical frequency of $\omega \simeq m_\pi$ and has a
Gaussian distribution of width $R_\perp$ in coordinate space,  
one can see from Eq.\ (\ref{eq:yield3}) that the dilepton yield 
will be concentrated around an invariant mass of $M\simeq 2 m_{\pi}$.
The width of this distribution will be of the order of $1/R_\perp$
in invariant mass as well as in the three-momentum $\bold{q}$. 

Within this schematic model, 
it is possible to take approximate account of expansion
by subjecting the \DCC\ field to a boost-invariant Bjorken expansion.
The \DCC\ field depends then only on the proper time $\tau=\sqrt{t^2-z^2}$
and, in the linear sigma model, its equation of motion becomes
\begin{equation}
  \ddot{\phi}+\frac{1}{\tau}\dot{\phi}+\lambda (\phi^2-v^2) \phi
  =H\hat{\sigma} \; . \label{eq:eom}
\end{equation}
The initial values of $\phi$ and $\dot\phi$
are sampled from normal distributions with suitable width parameters.
The corresponding energy-momentum tensor is then relatively simple,
\begin{equation}
  T^{\mu\nu}=\epsilon_K (2u^\mu u^\nu -g^{\mu\nu})
  +\epsilon_V g^{\mu\nu} ,
\end{equation}
where the four-velocity is $u^\mu=\tilde{x}^\mu/\tau$
with $\tilde{x_\mu}\equiv(t,0,0,z)$.
The contribution to the energy density from the time dependence of the field is
$\epsilon_K=\frac{1}{2}\dot{\phi}^2$,
while the contribution from the interaction is
$\epsilon_V=\frac{\lambda}{4}(\phi^2-v^2)^2-H\sigma-\epsilon_{\rm gs}$,
with the vacuum energy being
$\epsilon_{\rm gs}=\frac{\lambda}{4}(f_\pi^2-v^2)^2-Hf_\pi$.
The expansion causes the energy density to drop steadily,
\begin{equation}
  \frac{\partial}{\partial\tau}(\epsilon_K+\epsilon_V)
  =-2\frac{\epsilon_K}{\tau} \;\; ,
\end{equation}
and a simple power behavior is quickly approached,
$\epsilon_K+\epsilon_V\leadsto\epsilon_0\tau_0/\tau$.
The coefficient $\epsilon_0$ then provides a convenient means of
characterizing the particular solution.
The limiting $1/\tau$ behavior of the energy density
is a general characteristic of the boost invariant scenario \cite{JR:PRL}
and we therefore assume that the energy density of the thermal gas
drops in the same manner.
This is accomplished by using a time-dependent temperature for the gas,
$T=(\tau_0/\tau)^{1/4}T_0$.

We can take approximate account of the finite transverse size of a \DCC\ domain
by giving the chiral fields a common profile
(such that the orientation of the pion field in the Cartesian space
only depends on the proper time),
\begin{equation}
  g( x_\perp)=\exp(- x_\perp^2/2R_\perp^2)\ . \label{eq:prof}
\end{equation}
This transverse profile is simply an ansatz
and does not follow from nor is subjected to the equation of motion.
Such a simple Gaussian form was suggested by numerical simulations \cite{ahw}
and it suffices for our present purpose.
The   electromagnetic current density then becomes \cite{zhxw96}
\begin{equation}
j_\mu(x)\ =\ v_3\ f_\pi^2 \frac{\tilde{x}_\mu}{\tau^2}\ g^2( x_\perp)\ =\
v_3\ f_\pi^2 \frac{\tilde{x}_\mu}{\tau^2}\ \exp(- x_\perp^2/R_\perp^2)\ .
\end{equation}
where the dimensionless normalization coefficient $v_3$
is determined by the specific initial condition.

The longitudinal expansion causes the current density to decrease
as the proper time grows.
As a consequence,
the contribution to the dilepton yield
from the coherent dilepton emission remains finite.
Following Ref.~\cite{zhxw96},
and averaging over the ensemble of initial conditions,
we find 
\begin{equation}
  \frac{dN^{(0)}_{\ell^+\ell^-}}{dy_q dq_\perp^2 dM}\ =\
    \frac{\alpha^2}{24}\pi^2R_\perp^4\ {\rm e}^{- q_\perp^2 R_\perp^2/4}\
      f_\pi^4\ \prec v_3^2\succ\
    \frac{q_\perp^2}{M^3M_\perp^2}[J_0^2(M_\perp\tau_0)
    +N_0^2(M_\perp\tau_0)] \ , \label{eq:yield11}
\end{equation}
where the transverse mass is
$M_\perp=\sqrt{M^2+ q_\perp^2}$ (with $M^2=q^2$)
and $J_0$ and $N_0$ are Bessel functions. 
This differential yield is independent of the dilepton rapidity
because of the boost invariance of the \DCC\ field,
and so it depends only on the initial value of the current,
as determined by the ensemble average $\prec v_3^2\succ$.
It may also be noted that the coherent emission vanishes
as the transverse momentum of the dilepton photon approaches zero.
This further diminishes the importance of coherent emission
compared to the incoherent emission at low transverse momenta
where the effect of the \DCC\ is expected to be largest.

In order to calculate the incoherent dilepton production by the \DCC\ field,
we need to perform the Fourier transformation of pion field
$\bold{\pi}(x)=\bold{\pi}(\tau) g( x_\perp)$.
We find 
\begin{equation}\label{eq:pi_p}
  \pi_i(\pm p)= 
\left\{ \begin{array}{ll}
\pi G( p_\perp)\int_{\tau_0}^\infty \tau d\tau\
        \pi_i(\tau) \, [-N_0(M_\perp\tau) \mp i J_0(M_\perp\tau)]\ ,
  \;\;\;  &   \,\,  p_0 \, \geq |p_\parallel| \\ 
  2G( p_\perp) \int_{\tau_0}^\infty \tau d\tau\
        \pi_i(\tau) \, K_0(M_\perp\tau)\ ,
  \;\;\;  &     |p_0| \leq |p_\parallel|
\end{array}     \right.
\end{equation}
where $M_\perp=\sqrt{|p_0^2-p_\parallel^2|}$ and
$G( q_\perp)=\pi R_\perp^2 \exp(- q_\perp^2 R_\perp^2/4)$.

One can then use 
$\langle \pi^\dagger(q)\pi(q)\rangle=
\frac{1}{2}[\langle|\pi_1(q)|^2\rangle+\langle|\pi_2(q)|^2\rangle]$
in Eq.~(\ref{eq:yield3}) to calculate the dilepton yield
from the interaction of the thermal pions with the \DCC\ field,
\begin{eqnarray}
  \frac{dN^{(1)}_{\ell^+\ell^-}}{dy_qdq^2_\perp dM}
    &=&\frac{2\alpha^2}{3\pi^2}\frac{1}{M}
    \int\frac{d^3k}{2\omega_k (2\pi)^3}
    \left[\frac{(k\cdot q)^2}{M^2}-m_\pi^2\right] \nonumber \\
    & & \left[ (1+n^+_k) \langle\pi^\dagger(-q-k)\pi(-q-k)\rangle_{DCC}
       \right.
\nonumber \\
   &+&  (1+n^-_k) \langle\pi(q+k)\pi^\dagger(q+k)\rangle_{DCC}  
\nonumber \\
    &+& \left. n^+_k \langle\pi(q-k)\pi^\dagger(q-k)\rangle_{DCC}
    +n^-_k \langle\pi^\dagger(k-q)\pi(k-q)\rangle_{DCC} \right].
    \label{eq:yield31}
\end{eqnarray}
Due to the boost invariance,
we need only consider dileptons with zero rapidity
so that  $q=(M_\perp,{\bf q}_\perp,0)$.
Given the simple Gaussian transverse profile of the pion field,
one can carry out the azimuthal integral,
and the remaining two-dimensional integral
can then be evaluated numerically
by use of the Fourier transforms in Eq.\ (\ref{eq:pi_p}).

In the numerical evaluation of the dilepton production from the \DCC\ field,
we have employed the following rather conventional parameter values,
$\lambda=19.97$, $v=87.4$ MeV, $f_\pi=92.5$ MeV, $m_\pi=135$ MeV,
with which $H=f_\pi m_\pi^2=(119\ {\rm MeV})^3$ and
$m_\sigma=\sqrt{2\lambda f_\pi^2+m_\pi^2}=600$ MeV.
We first calculate the Fourier transformation of the pion fields
according to Eq.~(\ref{eq:pi_p}) for the solution of the equation of motion.
We then calculate $\langle \pi^\dagger(q)\pi(q)\rangle$
averaging over the initial configurations which then is used for
the numerical evaluation of the dilepton yield in Eq.~(\ref{eq:yield31}).
The ensemble average $\prec v_3^2\succ$ is also extracted
and used in the calculation of the coherent emission yield,
Eq.~(\ref{eq:yield11}).

\begin{figure}[htb]
\setlength{\epsfxsize=0.7\textwidth}
\centerline{\hspace{1cm}
\epsffile{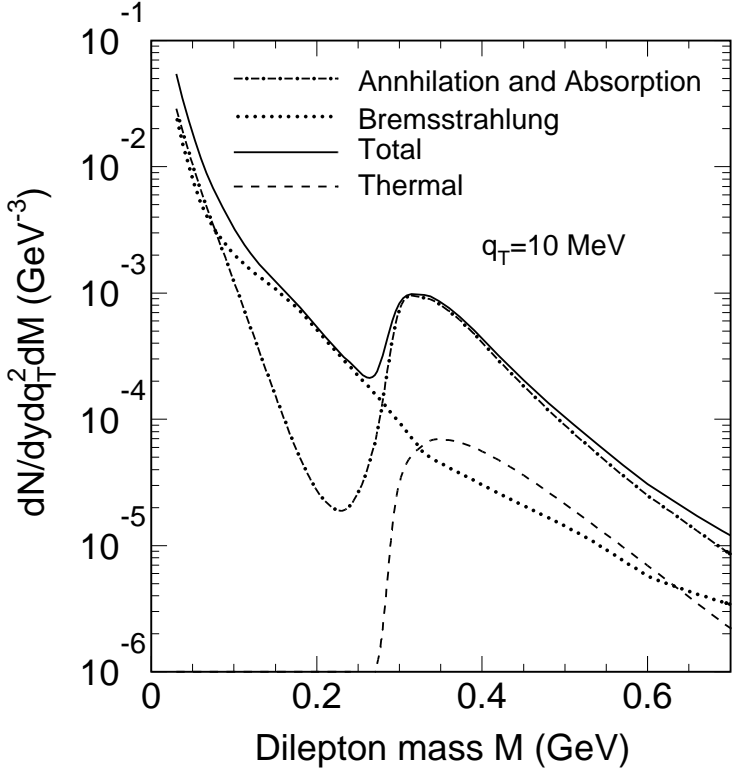}}
\caption{Dilepton spectrum from the \DCC\ field,
both bremsstrahlung (dotted),
annihilation and absorption (dot-dashed), and their sum (solid),
as well as the thermal emission (dashed). 
The initial temperature of the thermal environment is
$T_0$=145~MeV, while the characteristic energy density
carried by the \DCC\ field is $\epsilon_0$=58~MeV/fm$^3$.}
\label{fig:sch}
\end{figure}

Figure \ref{fig:sch} shows the differential dilepton yield from a total 
space-time volume\\
${\frac{1}{2}\pi R_\perp^2 (\tau_{\rm max}^2-\tau_0^2)}$
due to incoherent emission from the \DCC\ field
together with the corresponding average thermal rate.
We have used the following parameter values,
$T_0$=145~MeV, $R_\perp$=2~fm, $\tau_0$=1~fm, $\tau_{\rm max}$=102.4~fm,
and the characteristic energy density carried by the \DCC\ field was 
$\epsilon_0$=58~MeV/fm$^3$,
which is the same as the thermal pion energy density
at the employed temperature $T_0$.
We have also separated the incoherent dilepton rate into bremsstrahlung and 
annihilation (absorption) part, since they are caused by very different 
processes and should have different features in the invariant mass spectrum.

We find that the coherent emission rate in Eq.~(\ref{eq:yield11}) is 
negligible as compared to the incoherent emission.
This is related to the fact that the coherent emission in our scenario
depends only on the third component of the conserved isovector current.
Because the energy density carried by the isovector
current decreases much faster than that carried by 
other degrees of freedom,
the associated dilepton emission is thus also less important.
However, the incoherent dilepton emission is very significant
in comparison with the thermal production. The spectrum from
the pion annihilation and absorption by the \DCC\ field (dot-dashed)
has a structure manifest of two components in the second two terms
in Eq.~(\ref{eq:yield31}), depending on the energy flow.
In the contribution from the annihilation with the \DCC\ field,
energy flows out of the \DCC\ field, therefore the dilepton
spectrum has a threshold at $M=2m_\pi$, if the \DCC\ pion field
oscillate with a minimum frequency of $m_\pi$. If the energy flows
into the \DCC\ field, dileptons are then emitted when thermal pions
are absorbed by the \DCC\ field.
The dilepton spectrum from these absorption processes has no threshold
and dominates the annihilation and absorption spectrum at small invariant
masses $M<m_\pi$, as seen in Fig.~\ref{fig:sch}.

\begin{figure}[htb]
\setlength{\epsfxsize=0.7\textwidth}
\centerline{\hspace{1cm}
\epsffile{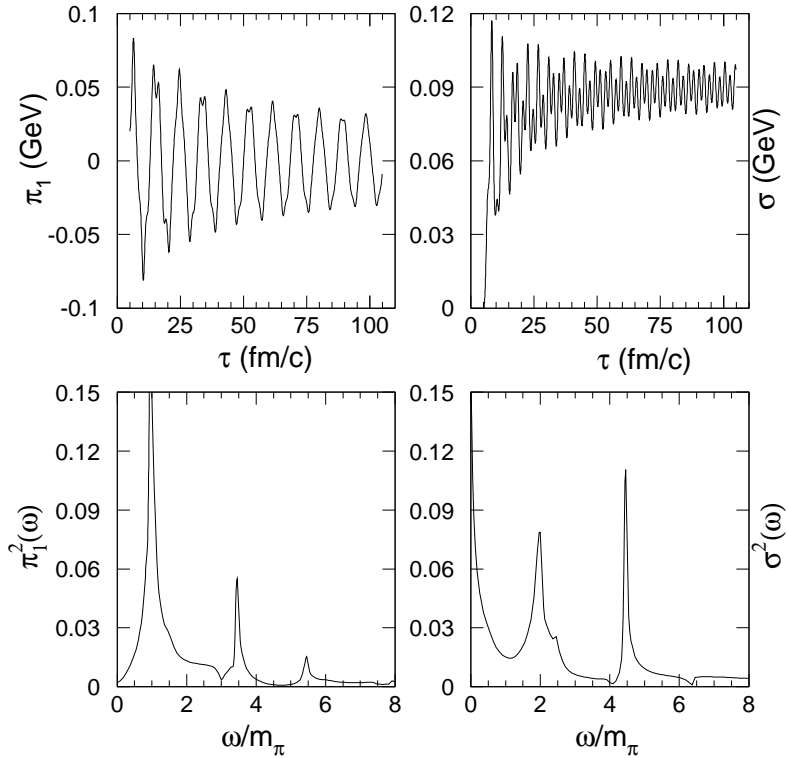}}
\caption{The time evolution of the $\pi_1$ and $\sigma$ \DCC\ field
and their corresponding Fourier spectra.}
\label{fig:para}
\end{figure}

Shown as the dotted line in Fig.~\ref{fig:sch} is the dilepton spectrum
due to pion bremsstrahlung from the \DCC\ field,
the first two terms in Eq.~(\ref{eq:yield31}).
Since the emitted pions have at least a minimum energy of $m_\pi$,
the \DCC\ field must have some higher frequencies
in order to emit a thermal pion plus a pair of leptons.
This means that the \DCC\ must have some quasi-particle modes
whose masses are larger than the pion mass $m_\pi$.
To illustrate this, we show in Fig.~\ref{fig:para}
both the pion and sigma fields as functions of time
and their corresponding Fourier spectra.
As one can clearly see,
besides the normal pion mode with mass $m_\pi$
(this is the frequency of the fields for the zero-momentum mode
we are considering),
there are others resonances with higher masses.
While the resonances at $\simeq 3.5 m_\pi$ and $ \simeq 5.5 m_\pi$ are 
clearly visible,
the higher ones cannot be seen here
due to their small amplitudes and the limited resolution
due to the finite time interval used in our numerical calculation.
These quasi-particle modes are normally referred to as parametric resonances
in a non-linear and strongly coupled system.
We will not elaborate on the interesting physics
associated with the parametric resonances,
but just point out that they play an important role
in the preheating of the early universe \cite{kls}
and the amplification of the long-wavelength
pion mode following an extremely nonequilibrium initial 
condition \cite{bvhs,smbm}
With these parametric resonances in mind,
we can readily understand the spectrum of the dilepton from
the bremsstrahlung processes: dileptons are produced via the
transitions of the high resonances to the normal pion mode.
As we increase the invariant mass of the dilepton,
some of the transitions are gradually turned off
as the energy of the dilepton becomes larger than the mass differences
between the higher resonances and the normal pion mass $m_\pi$.
This is why there is a slight oscillation
in the bremsstrahlung dilepton spectrum.
The oscillation vanishes when the invariant mass reaches
the highest parametric resonance that the numerical solution can resolve.
This is also the case when we increase the transverse momentum of the dilepton.

In Fig.~\ref{fig:sch} is also shown the sum of the different contributions
to dilepton spectrum from the \DCC\ field (solid curve)
as well as the contribution from annihilation in the cooling pion gas.
In general, one can see that the incoherent dilepton production
below and near $2m_\pi$ threshold region is significantly larger
than the thermal production.
As we already pointed out,
because of the finite spatial size of \DCC\ field in the transverse direction,
dileptons from the \DCC\ field exhibit a much faster decrease
with the transverse momentum $q_\perp$ than those due to thermal production. 


\section{Summary}

We have calculated the production of dileptons from disoriented chiral
condensates using a quantal mean-field as well as a semi-classical 
treatment for
the time evolution in the linear sigma model. We have compared the
dilepton spectra obtained when using so called quench initial conditions, 
which lead to a strong enhancement of long wave length pion modes (DCC), with
those obtained from thermal initial conditions. Compared to the thermal
spectrum the quench initial conditions lead to a strong enhancement (factor
20 - 100 depending on the model) at an invariant mass of about $M
\simeq 2
m_\pi$. This enhancement is confined to dilepton momenta  of 
$q \leq 300 - 500 \, \rm MeV$ 
and also rather narrow in invariant mass. Based on the 
analysis in a schematic
model we find that the observed
enhancement at $M \simeq 2 m_\pi$ is due to the annihilation of thermal
pions with those from the DCC.  

We furthermore found some evidence for an enhancement at low invariant masses, 
$M <  m_\pi$, which, however, we could not unambiguously establish due to
numerical backgrounds. In the schematic model this enhancement is due to
bremsstrahlung-type processes involving the absorption or emission of pions by
and from the DCC. The latter reflects the rich dynamical structure of the DCC
field. The spectral distribution of the pion field shows distinct peaks not
only close to the pion mass but also at higher frequencies. This phenomenon is
related to the so-called parametric resonances and is responsible for a
comparatively strong dilepton yield as a result of pion emission processes. 

Within the schematic model, 
we could also address the more realistic scenario of
a longitudinal expansion. We find that the above enhancement remains also in this
case. 

Both dynamical solutions, which are based on distinct approximations to
the linear sigma model, predict qualitatively the same enhancement:
A large bump at $M \simeq 2 m_\pi$
as well as an enhancement at low invariant masses.
Overall,
the quantal mean-field model seems to yield a larger enhancement
than the semi-classical treatment.
This quantitative difference may be due to several factors.
First, the Bose enhancement factors are absent in the semi-classical treatment.
Second, the semi-classical treatment incorporates the mode mixing
resulting from the non-linear interaction and this mechanism
tends to reduce the number of pions in the DCC state.
Both of these features lead to a somewhat smaller signal.

As far as experimental observation of this enhancement is concerned, 
it will be  probably
very difficult to see the enhancement at low invariant masses, 
$M <  m_\pi$, because this region will be dominated by the Dalitz decay of
the $\pi^0$. The enhancement around $M  \simeq 2 m_\pi$ on the other hand
should be observable in principle. In this mass range the major competing
channel is the Dalitz decay of the $\eta$. From the analysis of CERN SPS data
(see e.g. \cite{KS96}) the $\eta$ Dalitz is at most a factor of five stronger
than the pion-annihilation channel. Therefore an enhancement of the pion
annihilation by a factor of ten or larger should be visible. However, due to
the background from false pairs it is very difficult to 
measure dileptons of small
momentum \cite{carroll} where the expected enhancement is located. 

In the present work we have ignored expansion,
with the exception of the schematic model.
This will have to be taken into account in future work for a
more realistic description of a heavy-ion collision.
Morevoer, expansion provides one possible justification
of the quench initial conditions \cite{JR:PRL}.
Certainly in a given effective theory, a consistent treatment should 
generate the quench conditions dynamically starting from a more or less
thermal state at temperatures above $T_c$. The resulting dilepton spectrum
from such a calculation should then provide a more realistic picture of
possible enhancements to be expected in heavy ion experiments. 
In addition such a calculation needs to account for additional  
dissipative processes which are not included in the linear sigma model, such 
as the scattering with vector mesons.

However, too little is known about the physics that reigns close to the chiral
phase transition  to allow us to give a final theoretical answer about the 
existence of DCC states in relativistic heavy ion collisions. 
Therefore, experiments measuring unique observables,
such as the one addressed here,
will be needed to determine that issue and its implication
for the chiral phase transition in matter.

\section*{Acknowledgments}
We would like to thank J.D.~Bjorken, J.-P.~Blaizot, D.~Boyanovsky,
F. Cooper, A. Kovner, E. Mottola, and Z.~Huang for stimulating discussions.
This work was supported by the Director, Office of Energy Research,
Office of High Energy and Nuclear Physics, Divisions of High Energy Physics
and Nuclear Physics of the U.S. Department of Energy under Contract
No. DE-AC03-76SF00098.


\end{document}